\documentclass[twocolumn,superscriptaddress,showpacs,prl]{revtex4}
\usepackage{epsfig}
\usepackage{delarray}
\usepackage{amsmath, amssymb}
\usepackage{bm}

\newcommand{\ket}[1]{| \, #1 \rangle}
\newcommand{\bra}[1]{ \langle #1 \,  |}

\begin{document}
\title{Demonstration of local expansion toward large-scale entangled webs}
\author{Toshiyuki Tashima}
\affiliation{Graduate School of Engineering Science, Osaka University, Toyonaka, Osaka 560-8531, Japan}
\author{Tsuyoshi Kitano}
\affiliation{Graduate School of Engineering Science, Osaka University, Toyonaka, Osaka 560-8531, Japan}
\author{\c{ S}ahin Kaya \"Ozdemir}
\affiliation{Department of Electrical and Systems Engineering, Washington University, St. Louis, MO 63130 USA}
\author{Takashi Yamamoto}
\affiliation{Graduate School of Engineering Science, Osaka University, Toyonaka, Osaka 560-8531, Japan}
\author{Masato Koashi}
\affiliation{Graduate School of Engineering Science, Osaka University, Toyonaka, Osaka 560-8531, Japan}
\author{Nobuyuki Imoto}
\affiliation{Graduate School of Engineering Science, Osaka University, Toyonaka, Osaka 560-8531, Japan}

\begin{abstract}
We demonstrate an optical gate
that increases the size of polarization-based W states by accessing only
one of the qubits. Using this gate, we have generated three-photon and
 four-photon W states with fidelities $0.836\pm 0.042 $ and $0.784\pm 0.028$,
 respectively. We also confirmed the existence of pairwise entanglement in
every pair of qubits, including the one that was left untouched by
the gate. The gate is applicable to any size of W states and hence is a
universal tool for expanding entanglement.
\end{abstract}
\maketitle
For the past two decades much effort has been devoted to the study of
entanglement in order to grasp its nature \cite {s0} and to use it as a resource
for various quantum information tasks such as quantum key distribution
\cite {s1}, quantum metrology \cite {s2}, and quantum computing \cite {s3}. While entanglement between two quantum systems is well-understood,
entanglement among three or more quantum systems still requires
intense research effort. Even when we limit the constituent systems
to the simplest ones, qubits, we still encounter many nonequivalent
classes of entanglement which differ in the structure of how the
qubits are correlated. In a Greenberger-Horne-Zeilinger (GHZ) state, the entanglement is sustained
by all of the qubits \cite {s8}, in the sense that removal of any one qubit
completely disentangles the rest. Qubits in W states are entangled in
the completely opposite way, where entanglement is compartmentalized
such that entanglement between any pair of qubits survives after
discarding the rest of the qubits \cite {s8, s9, s10}. Cluster states
are halfway between the above two classes: Entanglement in the $N$-qubit
linear cluster states is sustained by at least half of the qubits; that
is, accessing $N/2$ qubits is enough to destroy the entanglement
completely \cite {s11}. 

Recently, expansion gates have been introduced for preparing large-scale
multipartite photonic entanglement \cite {s13, s14, s15, s16, s17, s18,
s19}.  In this approach, multipartite entangled states of a certain
class are grown from a small seed by locally adding one or more qubits
at a single site while retaining the structure of the desired entanglement class.

\begin{figure}[tb]
\begin{center}
\includegraphics[scale=1]{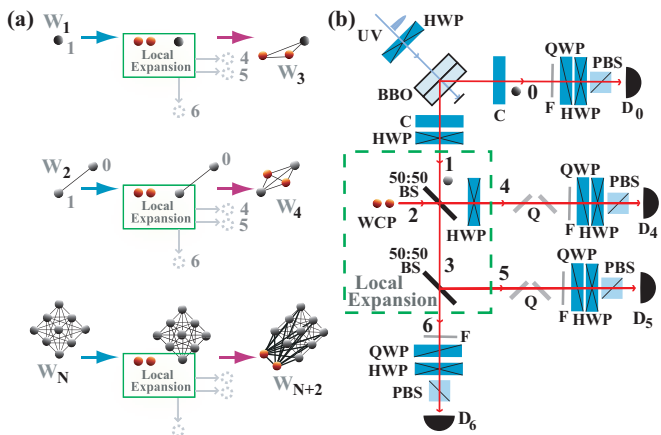}
\caption{\textbf{(a)} Concept of
 local state expansion. \textbf{(b)} Experimental
 setup. BBO: $\beta$-barium borate crystals (thickness 1.5mm + 1.5mm). HWP: half-wave plate. QWP:
 quarter-wave plate. UV: ultraviolet pulse.
 BS: beamsplitter. IF: narrowband interference filter (wavelength 790nm, bandwidth 2.7nm).Q: quartz
 crystals. C: BBO (thickness 1.65mm) used to compensate walk-off effects. Although the ideal
 operation of the gate requires two-photon Fock state in mode 2, in the
 experiments the two photons of mode 2 are provided by a weak coherent
 pulse (WCP). All the detectors ($\rm D_{0,4,5,6}$) are silicon avalanche photodiodes placed after single-mode optical fibers. \label {fig:1s}}
\end{center}
\end{figure}

It is well known that GHZ states and cluster states are, in
principle, deterministically expanded by applying a
controlled-unitary gate between one of the entangled qubits and a
fresh qubit to be added \cite {s13, s14}. Probabilistic implementations with
linear optics based on quantum parity check have also been
demonstrated \cite {s15, s16, s17}. The expansion of W states is much more
complicated for several reasons. First of all, when we expand an
$N$-qubit W state by accessing only one of the $N$ qubits, the
marginal state of the remaining $N-1$ qubits must be changed to the
proper state, which depends on the size of the expanded larger W
state. Thus, no unitary gates can be used for the expansion of W
states, even in principle. Another complication arises from the
difference in the structure of multipartite entanglement. As
mentioned above, each pair of qubits in a W state sustain its own
entanglement independently, resulting in a weblike structure of
bonding among qubits \cite {s9, s10}. In order to expand an
$N$-qubit W state while retaining such a structure, the newly added
qubits should not only get entangled with the accessed qubit in the
initial W state, but should also form independent pairwise entanglement with
each of the untouched $N-1$ qubits [see Fig.\ \ref {fig:1s}(a) ]. 

In this Letter, we demonstrate an
experimental implementation of this expansion task with a
surprisingly simple gate. The gate shown in the dotted box of Fig.\ \ref {fig:1s}(b) is essentially composed of just
two half beamsplitters with two-photon Fock states as a source of
fresh qubits \cite {s18}. The
gate involves three photons in total, including one photon from input mode 1 and two photons in H polarization in mode 2. The
successful operation of the gate is defined to be the case where one
photon emerges at each of the three output modes 4, 5, and 6. This
gate can be used for the expansion of polarization-entangled W states
written in the form 
\begin{eqnarray}
\ket {{\rm W}_N}= \frac{1}{\sqrt{N}}\sum_{j=1}^{N} a^\dagger_{j_{\rm
V}}\left(\prod_{i\neq j}
a^\dagger_{i_{\rm H}}\right)\ket {vac} \label{eq:1}
\end{eqnarray}
where $\ket{vac}$ denotes the vacuum state for all modes and
$a^\dagger_{i_{\rm V(H)}}$ is the creation operator of a
V (H)-polarized photon in mode $i$. The gate is size independent; namely, the same
gate is applicable to the expansion of an $N$-qubit polarization-based W state of any size
$N$ to produce an $N+2$-qubit polarization-based  W state. In this case,
the photon in mode 1 is provided by the $N$-photon polarization-based W
state. Here we present an experimental demonstration of the gate to
prepare three- and four-photon polarization-based W states corresponding to $N=1$ and $N=2$, respectively.

In the following, we will briefly explain the working principle of the expansion gate depicted in the dotted box in Fig.\ \ref
{fig:1s}(b). Let us first consider the case where the input photon is V polarized,
namely, state
$\ket {1_{\rm V}}_1\equiv \hat{a}^{\dag}_{1_{\rm V}}\ket{vac}_{1}$,
where $\ket{vac}_{1}$ stands for the vacuum for mode $1$. This photon
is distinguishable from the two photons of  mode 2 in state $\ket {2_{\rm H}}_2$ by polarization.
Hence the classical particle picture is applicable, namely, V, H, and
H photons emerge at modes 4, 5, and 6, respectively, with a probability
of $1/16$. The successful operation involves the other two cases, $\rm HVH$
and $\rm HHV$, too. The relevant state transformation for the two beam
splitters (BSs) is thus written as
\begin{eqnarray}
\ket {1_{\rm V}}_1&\rightarrow&\frac {1}{4}[-\ket {1_{\rm V}}_4\ket
{1_{\rm H}}_{5}\ket
{1_{\rm H}}_{6}+\ket {1_{\rm H}}_4\ket {1_{\rm
V}}_{5}\ket
{1_{\rm H}}_{6}\nonumber\\&&+\ket {1_{\rm H}}_4\ket {1_{\rm
H}}_{5}\ket
{1_{\rm V}}_{6}],\label{eq:2}
\end{eqnarray}
where the minus sign appears when the photon in mode 1 is reflected
toward mode 4. The half wave-plate (HWP) in mode 4 is used to
compensate this sign by introducing a $\pi$ phase shift on the
V polarization. The transformation by the entire gate is thus given by 
\begin{eqnarray}
\ket {1_{\rm V}}_1&\rightarrow&\frac {1}{4}[\ket {1_{\rm V}}_4\ket
{1_{\rm H}}_{5}\ket
{1_{\rm H}}_{6}+\ket {1_{\rm H}}_4\ket {1_{\rm
V}}_{5}\ket
{1_{\rm H}}_{6}\nonumber\\&+&\ket {1_{\rm H}}_4\ket {1_{\rm
H}}_{5}\ket
{1_{\rm V}}_{6}]=\frac {\sqrt{3}}{4}\ket {\rm W_3},\label{eq:3}
\end{eqnarray}
which means that a V-polarized photon is expanded to a three-photon W
state with a probability of $3/16$.

When the input photon in mode 1 has H polarization, the
transformation is simply given by changing $\rm V$ to $\rm  H$ in Eq.\
(\ref{eq:2}).
Since the HWP has no effect, this leads to
\begin{eqnarray}
\ket {1_{\rm H}}_1&\rightarrow&\frac
{1}{4}\ket {1_{\rm H}}_4\ket {1_{\rm
H}}_{5}\ket {1_{\rm H}}_6.\label{eq:4}
\end{eqnarray}
Here the success probability is reduced by a factor of 3, due to the
destructive interference caused by the indistinguishability of photons.

The transformation for a general input is now calculated from Eqs.\ (\ref{eq:3}) and (\ref{eq:4}).
When the input is one photon of the bipartite entangled state $\rm
\ket {\rm W_2}\equiv
(\ket{1_{\rm H}}_0\ket{1_{\rm V}}_{1}+\ket{1_{\rm V}}_0\ket{1_{\rm
H}}_{1})/\sqrt{2}$, this gate performs the transformation $\ket {\rm
W_2} \rightarrow (1/ \sqrt 8)~\ket {\rm W_4}$
resulting in a four-partite W state with a probability of $1/8$. Note that the photon in mode 0 of $\ket {\rm W_2}$ is untouched by this
gate. When the input photon in mode 1 is provided by an $N$-photon W
state, the gate performs the transformation $\ket {{\rm W}_N}
\to \sqrt {(N+2)/16N}~\ket{{\rm
W}_{N+2}}$. Hence the gate is applicable to any $N$, with a success probability
of $(N+2)/(16N)$. 

Our experimental setup designed for the realization of
this gate is shown in Fig.\ \ref {fig:1s}(b). The light pulses from a mode-locked Ti:sapphire
laser (wavelength 790nm, pulse width 90fs, repetition rate 82MHz) are divided into two unequal parts by a tilted glass plate. The weak portion is used to prepare an H-polarized weak coherent pulse
(WCP) with adjustable mean photon number $\nu \ll 1$. With a probability
of $\sim \nu^2/2$, this pulse includes two photons, which are used as
the ancillary state $\ket {\rm 2_H}$. The
 strong portion goes to a second harmonic generator to prepare
 the ultraviolet (UV) pulse used for photon pair generation by a
 spontaneous parametric down conversion (SPDC) in Type I phase
 matched $\beta$-barium borate (BBO) crystals stacked
 together with their optical axes orthogonal to each other \cite
 {s29}. The photon pair
 generation rate $\gamma$ is adjusted such that $\gamma\ll\nu\ll 1$ is
 satisfied. This ensures that the events with the WCP having one photon make
little contribution to the coincidence detection.

As a preliminary experiment, we made sure that a single photon from SPDC
was in a well-matched mode with the WCP when they were overlapped
at the first 50:50 BS. V-polarized UV pump pulses of an average power 23 mW are used for the
SPDC, and detection of one H photon in mode 0 prepares an H-polarized
single photon in mode 1. The H-polarized WCP in mode 2 is set to have
$\nu = 0.03$. The three-fold coincidences at modes 0, 4, and 5 were
recorded while varying the delay at mode 2 using a motorized
stage. A Hong-Ou-Mandel dip with visibility 0.85 at zero delay was observed [see Fig.\ 2]. This indicates a good overlap
between the two modes. 

\begin{figure}[tb]
\begin{center}
\includegraphics[scale=1]{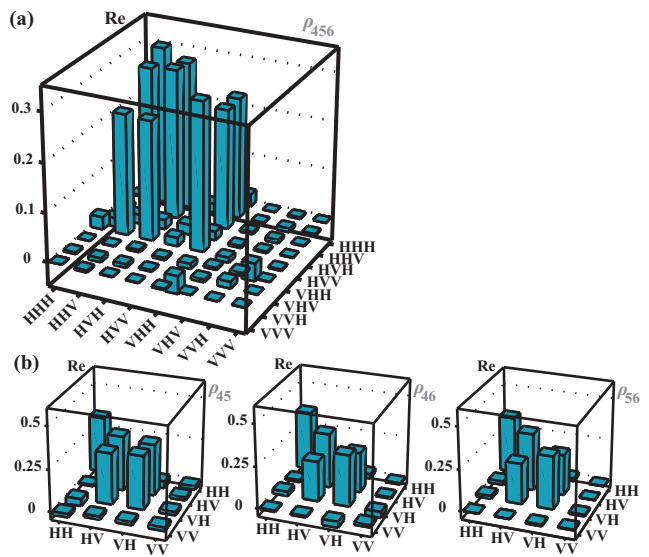}
\caption{Results of the expansion of $\ket {\rm W_1}$ to a three-photon polarization-entangled
 W state $\ket {\rm W_3}$. \textbf{(a)} The real part of the reconstructed
 density matrix of the final state $\rho_{\rm 456}$. \textbf{(b)} The real
 parts of the reconstructed reduced density matrices $\rho_{\rm 45}$,
 $\rho_{\rm 46}$, and $\rho_{\rm 56}$.\label {fig:2s}}
\end{center}
\end{figure}

As a demonstration of our expansion gate, we first fed a V-polarized
single photon ($\ket {{\rm W}_1}$) to the gate to produce $\ket {{\rm W}_3}$. The V-polarized single
photon is prepared by rotating the polarization of the single photon
prepared in the preliminary experiment by $\pi/2$ using the HWP at mode 1. The
successful operation of the gate is post-selected by the four-fold
coincidence detection at modes 4, 5, 6, and 0. We set $\nu = 0.3$ for the WCP
and an average power of 75 mW for the UV pump, and reconstruct the
density matrix $\rho_{456}$ of the three photons at modes 4, 5, and 6
using the iterative maximum likelihood method (IMLM) from the polarization-correlation measurements
on 64 different settings formed by combinations
of the single photon projections to $\rm \ket {H}$, $\rm \ket {V}$, $\rm
\ket {D} = (\ket {H}+\ket {V})/\sqrt 2$, $\rm \ket {R} = (\ket {H}-i\ket
{V})/\sqrt 2$, and $\rm \ket {L} = (\ket {H}+i\ket {V})/\sqrt 2$ \cite {s21, s22}. The coincidences were recorded for an acquisition time of 5220s at each tomographic setting with a typical
fourfold coincidence rate of $\sim 0.02$ counts/s. The reconstructed
density matrix $\rho_{456}$ shown in Fig.\ \ref {fig:2s}(a) carries a similar structure as the ideal $\ket {{\rm W}_3}$
which consists of only nine real and nonzero terms, namely, the diagonal
terms corresponding to $\ket {\rm HHV}$, $\ket {\rm HVH}$ and $\ket {\rm
VHH}$ and six off-diagonal elements corresponding to coherence among
these terms. Fidelity of the output state to $\ket {{\rm W}_3}$  is calculated as $F_{456}\equiv \bra {{\rm W}_3}\rho_{456}\ket
{{\rm W}_3}=0.836\pm 0.042$, and the expectation value of the entanglement
witness operator ${\cal W}_{\rm W} = \frac {2}{3}{\bf 1} - \ket {\rm
W_3}\bra {\rm W_3}$ is calculated as ${\rm Tr}({\cal  W}_{\rm
W}\rho_{\rm 456}) = -0.169\pm 0.042$, whose negativity proves that the prepared state
has genuine tripartite entanglement \cite {s23,s24}. In order to confirm the presence of the pairwise entanglement in the
 prepared W state, we
 calculated  the marginal density matrices of pairwise combinations
$\rho_{45}$, $\rho_{46}$ and $\rho_{56}$ from the reconstructed state
$\rho_{\rm 456}$ and depicted them in Fig.\ \ref {fig:2s}(b). We also
calculated the values of
entanglement of formation (EOF) as $\mathcal{E}(\rho_{45})=0.354\pm
0.070$, $\mathcal{E}(\rho_{46})=0.273\pm 0.065$ and
$\mathcal{E}(\rho_{56})=0.316\pm 0.074$, respectively \cite {s25}. All pairwise components of the output state $\rho_{\rm
456}$ enjoy entanglement, demonstrating that the state expansion gate
acts as an entangling gate and expands a single photon state into a tripartite W state. These results imply that the transformation given in Eq.\ (\ref{eq:3}) has been performed successfully in our experiments.
\begin{figure}[tb]
\begin{center}
\includegraphics[scale=1]{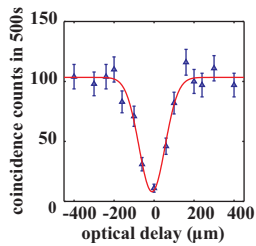}
\caption{Observed two-photon interference. The best fit to the
 experimental data is represented by the solid Gaussian curve (coherence
 length $l_c \simeq 144 \rm \mu m$ and visibility is $0.85$).\label {fig:1.1s}}
\end{center}
\end{figure}

\begin{figure}[tb]
\begin{center}
\includegraphics[scale=1]{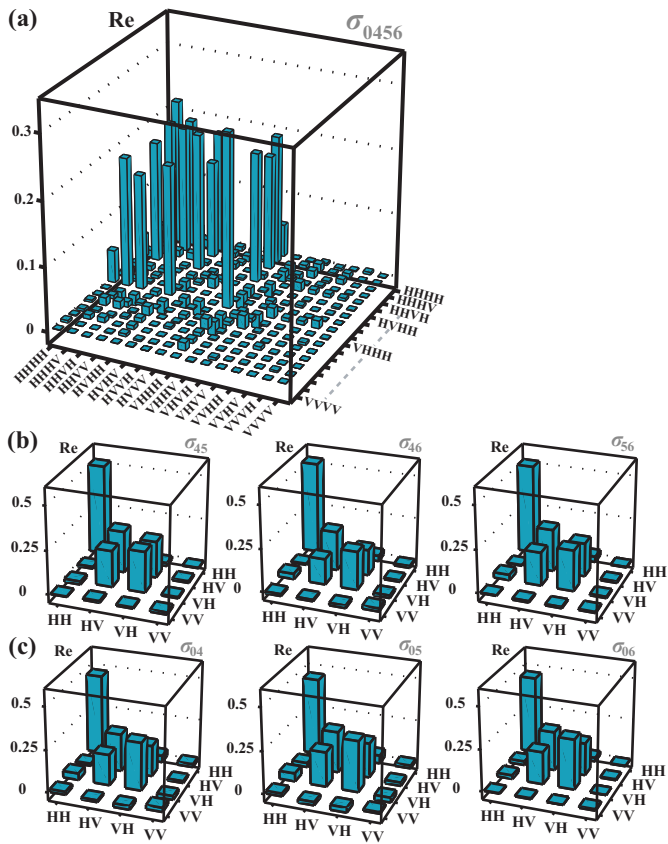}
\caption{Results of the expansion of $\ket {\rm W_2}$ to a four-photon
 polarization-entangled W state $\ket {\rm W_4}$. \textbf{(a)} The real part of the
 reconstructed density matrix of the final state $\sigma_{\rm 0456}$. \textbf{(b)} The real parts of the reconstructed reduced density matrices $\sigma_{\rm 45}$, $\sigma_{\rm 46}$, and $\sigma_{\rm 56}$, and \textbf{(c)} those of $\sigma_{\rm 04}$, $\sigma_{\rm 05}$, and $\sigma_{\rm 06}$.\label {fig:3s}}
\end{center}
\end{figure}

Next we proceed to the expansion of a bipartite entangled state ($\ket
 {{\rm W}_2}$)
 into $\ket {{\rm W}_4}$. By setting the
polarization of UV pulses (average power: 150mW) to diagonal polarization, we prepared the bipartite entangled state $\sigma_{01}$ and characterized it by reconstructing its density matrix by measuring polarization correlations in modes 0 and 6 using 16
different basis settings. Fidelity to a maximally entangled photon pair $\rm F_{01}\equiv \bra {{\rm W}_2}\sigma_{01}\ket
{{\rm W}_2}$ and the EOF of the prepared state are
calculated as $F_{01}=0.977\pm 0.005$ and $\mathcal{E}(\sigma_{01})=0.964\pm 0.013$, respectively,
confirming that the prepared state is a highly entangled photon
pair. Then we mixed the photon in mode 1 of $\sigma_{01}$ with the WCP ($\nu = 0.3$) and post-selected the successful events by fourfold
 coincidences in modes 0, 4, 5 and 6. The state $\sigma_{0456}$ of the
 four photons was reconstructed  using 256 different tomographic settings. The coincidences were recorded for an acquisition time of 4280s at each tomographic setting with a typical fourfold coincidence rate of $\sim
0.02$ counts/s. The density matrix of $\sigma_{0456}$ was reconstructed
 using the IMLM (Fig.\ \ref {fig:3s}(a)). The density matrix of an ideal
 $\ket {{\rm W}_4}$ state consists of 16 real and nonzero elements
 including twelve off-diagonal elements depicting the coherences among the four diagonal terms $\ket {\rm HHHV}$, $\ket {\rm HHVH}$, $\ket {\rm HVHH}$
and $\ket {\rm VHHH}$. A similar structure is clearly seen in the
 density matrix of the output state $\sigma_{0456}$ of the expansion gate. From the reconstructed density matrix, we calculated
the fidelity as $F_{0456} \equiv \bra {\rm W_4}\sigma_{0456}\ket {\rm W_4} =0.784\pm
0.028$. The entanglement witness calculated using the operator ${\cal W}_{\rm W} = \frac {3}{4}{\bf 1} - \ket {\rm
W_4}\bra {\rm W_4}$ has the ideal expectation value of $-1/4$ for an
ideal $\ket {\rm W_4}$ \cite {s24}. The presence of genuine four-partite entanglement in
$\sigma_{0456}$ is confirmed by the calculated expectation value of ${\rm
Tr}({\cal W}_{\rm W}\sigma_{\rm 0456}) = -0.034\pm 0.028$.

It is known that the larger the W state, the weaker the pairwise
entanglement, which makes it more challenging to observe experimentally. We have calculated two-qubit marginal density operators
for various combinations of the reconstructed density operator
$\sigma_{0456}$. Figure \ref {fig:3s}(b) shows the three combinations for the qubits
4, 5 and 6, which have been directly interacted at the gate. The
calculated values of EOF for these cases are $\mathcal{E}(\sigma_{45})=0.040\pm 0.022$, $\mathcal{E}(\sigma_{46})=0.167\pm 0.033$ and $\mathcal{E}(\sigma_{56})=0.133\pm 0.030$, which are all
positive. Figure \ref {fig:3s}(c) shows the two-qubit density operators involving
the photon in mode 0, which has been untouched by the gate. The calculated
values of the EOF are
all positive: $\mathcal{E}(\sigma_{04})=0.184\pm 0.037$, $\mathcal{E}(\sigma_{05})=0.072\pm 0.028$ and $\mathcal{E}(\sigma_{06})=0.146\pm 0.033$. Although experimentally
achieved EOF are smaller than the theoretical maximum of 0.35, they are
a conclusive sign of the presence of pairwise entanglement in every
pair, which is the crucial property of $\ket {\rm W_4}$. This result
also confirms that the transformations in Eqs.\ (\ref{eq:3}) and
(\ref{eq:4}) were carried out coherently in our experiment. 

If we interpret that the photon in mode 1 has come out in mode 4 after the
 interaction with the two photons in mode 2, we may say that the
 pairwise entanglement in $\sigma_{05}$ and $\sigma_{06}$ were newly created as a
 result of the gate operation, while the original strong entanglement $\mathcal{E}(\sigma_{01})=0.95\pm 0.02$ was reduced to $\mathcal{E}(\sigma_{04})=0.15\pm 0.03$. One
 may also argue that the three photons 4, 5, and 6 are
 indistinguishable, and hence none of them are entitled to be the exclusive
 descendant of the photon in mode 1. This picture gives rise to another interesting
 interpretation, in which the photon in mode 1 has been cloned into the three
 copies, 4, 5, and 6. In contrast to the conventional 1 $\to$ 3 optimal
 cloning in which the only aim is to copy the state (including
 correlations to a reference system) of the input qubit as good as
 possible \cite {s26}, our gate produces pairwise entanglement among the three
 output qubits, in addition to the conventional task. The gate is
 optimal in achieving both of the tasks, which comes from the optimality
 of the pairwise entanglement in $\ket {\rm W_4}$. Figures \ref {fig:3s}(b) and \ref {fig:3s}(c) show that both of
 the tasks are achieved at the same time in our experiment. 

In summary, we have shown that the size of a multipartite
entanglement can be expanded by operating a simple gate on one local
site, preserving the characteristic entanglement structure over the whole
system including the sites to which the gate has no access. We
demonstrated expansion of W states up to the size of four, and the same
gate is expected to be applicable to any size of W states. Thanks to the
simple structure,  the expansion gate can be easily miniaturized by
integration on recently developed silicon waveguide quantum circuits
\cite {s28}. We believe that the
demonstrated expansion gate has the potential to become an integral part
of any quantum optical toolbox aimed at the preparation, manipulation
and understanding of multipartite entangled states.

This work was supported by the Funding Program for World-Leading Innovative
     R\&D on Science and Technology (FIRST), MEXT Grant-in-Aid for Scientific
Research on Innovative Areas No. 20104003 and No. 21102008, JSPS Grant-in-Aid for
Scientific Research(C) No. 20540389, and MEXT Global COE Program.

\end{document}